# MULTISCALE MODELS FOR PEROVSKITE OPTIMISATION

Philippe. Baranek[1,2,*], James P. Connolly[3], Antoine Gissler[1,2,4], Philip Schulz[4], Michel Rérat[5] and Roberto Dovesi[6]

[1] EDF R&D, EDF Lab Paris-Saclay, Department SYSTEME, 7 boulevard Gaspard Monge, F-91120 Palaiseau, France
[2] IPVF, Institut Photovoltaïque d'Ile-de-France, 18 boulevard Thomas Gobert, F-91120 Palaiseau, France
[3] [1]GeePs, Group of Electrical Engineering Paris, CNRS, CentraleSupelec, Université Paris-Saclay, Sorbonne Université, 3&11 rue Joliot-Curie, Plateau de Moulon, 91192 Gif-sur-Yvette CEDEX, France
[4] École Polytechnique, IPVF, UMR 9006, CNRS, 18 boulevard Thomas Gobert, F-91120 Palaiseau, France
[5] Université de Pau et des Pays de l'Adour, E2S UPPA, CNRS, IPREM, 2 avenue du Président Pierre Angot, F-64053 Pau, France
[6] Accademia Delle Science di Torino, via Accademia delle Science 6, I-10123 Torino, Italy

ABSTRACT: This paper presents a multiscale approach to evaluate perovskite solar cell performance which determines material properties at the atomistic scale with first-principles calculations, and applies them in macro-scale device models. This work focuses on the MAPbI$_3$ (MA = CH$_3$NH$_3$) perovskite and how its phase transitions impact on its optical, electronic, and structural properties which are investigated at the first-principles level. The obtained data are coupled to a numerical drift-diffusion device model enabling evaluation of the performance of corresponding single junction devices. The first-principles simulation applies a hybrid exchange-correlation functional adapted to the studied family of compounds. Validation by available experimental data is presented from materials properties to device performance, justifying the use of the approach for predictive evaluation of existing and novel perovskites. The coupling between atomistic and device models is described in terms of a framework for exchange of optical, vibrational, and electronic parameters between the two scales. The result of this theoretical investigation is a methodology for designing and optimising perovskite materials for both cell performance and stability, the key obstacle in the societal implementation of these record-breaking new materials.

## 1 INTRODUCTION

Perovskite solar cells have progressed extremely rapidly from 3.8% in 2009 to 27.3% in September 2024. Tandem efficiencies have furthermore breached the single-junction Shockley-Queisser efficiency limit, reaching 34.6% in June 2024 (LONGI, certified) [1, 2]. While this rapid efficiency increase is unmatched by any other technology, it remains crippled by stability issues, obstacle for the industrial and societal application of these materials. State of the art perovskite absorber materials still suffer stability issues linked to temperature, to volatile organic cations for the organic case and its reactivity to the air moisture among other issues. Both air moisture and temperature induce phase transitions which degrade the performance and durability of perovskite solar cells (PSCs): the moisture leads to the appearance of a non-perovskite phase (the so-called δ black phase) which is optically inactive, while the temperature can lead to a rich sequence of phase transitions. Their impact concerns mainly the electronic properties and the domains and surface stabilities of the different compounds. A key element is stability implications of the effect of phase transitions on the nonlocal lattice distribution of organic moieties through the lattice

In this work, we focus on the phase transitions impacts on the cell efficiency. They are associated in particular to the existence of soft phonon modes which can locally generate phase instabilities. For both organic and inorganic perovskites, they are linked to the lattice and halide octahedra deformations. However, for the organic case, another factor has to be taken into account which is the nonlocal ordering of the organic moieties inside the lattice through the different phase transitions.

Li and co-workers [3, 4] showed that the inorganic-framework deformation depends on the orientation of the organic cation which directly influences the stability of the hybrid perovskites and deserves a multiscale approach to obtain a good description of their properties.

If we consider CH$_3$NH$_3$PbI$_3$ as a paradigmatic case from an experimental point of view, the difficulty in obtaining an accurate characterization of its phase transitions comes from the determination of the methalominium (CH$_3$NH$_3^+$, MA) atomic positions inside the PbX$_3$ lattice: since the measurements are mainly performed with X-Ray diffraction, the positions of the MA moiety are ill or not defined. Therefore, for the *Pm3m* cubic phase the commonly used assumption is to consider MA as an intrinsic chemical entity which lies in the center of the cubic cell. However, this is not consistent from a crystallographic point of view: For instance, since the MA point group is $C_{3v}$, the corresponding space groups is $C_{3v}$ *(R3m)* if the C–N bond is along the [1,1,1] direction of the cubic cell. Moreover, with this description MAPbI$_3$ is necessarily in a ferroelectric phase which might lead to a wrong characterization of its optoelectronic properties.

In this paper, using a theoretical multiscale approach, we illustrate how phase transitions can impact the performance of solar cells. This modelling couples atomistic scale first-principles calculations to device scale numerical models. The coupling between atomistic and device models is described in detail by presenting a framework for exchange of optical, and electronic parameters between the two scales.

This approach is based on a crystallographic description of MAPbI$_3$ which allows to take the MA ordering into account. We first describe this crystallographic model. At the first principles level, it is used to determine a hybrid exchange-correlation functional adapted to the $MAPbX_3$ (x = Cl, Br and I) family of perovskites. The evolution induced by the phase transitions on the electronic and dielectric properties of $MAPbI_3$ is then systematically investigated. The corresponding band gaps, electron affinities and dielectric responses serve as input data to the device model which integrates these data in the absorber of a standard perovskite solar cell design [5] which we will not detail here. This device model yields the corresponding solar cell performance allowing evaluation of the impact of materials configurations at the atomistic scale on device performance and stability.

---

[*]Corresponding author: philippe.baranek@edf.fr

## 2 METHODOLOGICAL ASPECTS

### 2.1 First-principles approach

We define a crystallographic structure allowing evaluation of the ordering of the MA moities inside the lattice through the different phase transitions (see figure 1).

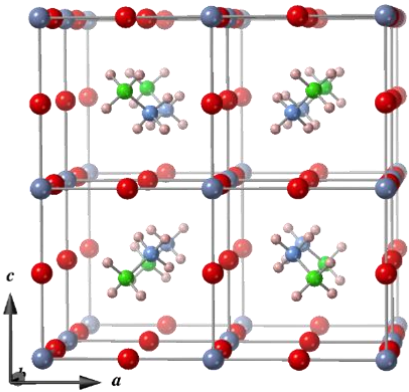

**Figure 1:** Used crystallographic structure of the *Pm3m* cubic phase of $MAPbI_3$.

First-principles calculations have been performed with the use of the CRYSTAL code [6, 7]. This program enables solution of both the Hartree–Fock (HF) and the Kohn–Sham (KS) systems of equations, combining them within a hybrid scheme. This work uses a hybrid exchange-correlation functional optimized to yield description of the structural, electronic, and dynamic properties of $MAPbX_3$ (X = Cl, Br and I) in good agreement with experiment, and has recently been used efficiently to study the influence of alkali metals on the properties of chalcopyrites, perovskites surface properties and the humidity-induced degradation products of halide perovskites [8-10]. In this work, the Hamiltonian (denoted as **PBEx**) combines 19% of HF exact exchange with the PBE exchange correlation functional [11]. It provides results consistent with a more homogeneous quantitive description of their properties than the most commonly used screened hybrid functional HSE [12] consistent with the most sophisticated methods based on the GW approximation: the obtained mean absolute average errors on the lattice parameters and band gaps of their different phases are 2 and 5 %, respectively, with respect to the available experimental data (as illustrated by the Table II for the cubic phase).

At the first-principles level, the changes induced by the phase transitions of the electronic, vibrational, and dielectric properties of each perovskite is systematically investigated. The resulting band gaps, work functions and dielectric responses serve as input data to the device model which yields the performance of solar cells.

### 2.2 Device model

The device scale numerical modelling is performed by on SILVACO's ATLAS simulator [13]. This uses the drift diffusion model, solving the current, continuity, and Poisson equations on a one to three dimension mesh. The full list of parameters identified are summarised in table I. The multiscale coupling consists of identifying device level parameters which can be provided by atomistic scale density functional theory materials models.

**Table I:** Full set of device scale drift-diffusion (DD) model inputs from atomistic scale density functional theory (DFT) level. This study uses a subset which are band parameters and optical functions.

| Parameter | Definition |
| --- | --- |
| $\tau_{SRH}$ | Electron and hole charge neutral and depletion layers Shockley-Read-Hall lifetimes |
| $\mu$ | Carrier mobility, majority and minority, electron and hole |
| $D_N$, $D_P$ | Hole and electron diffusion coefficients |
| $C_A$ | Auger coefficient |
| $\varepsilon$ | Permittivity related to complex refractive index |
| n, k | Real and imaginary refractive indices |
| $m_e^*$ | Electron and hole effective masses |
| $\chi$ | Electron affinity |
| $N_C$, $N_V$ | Band parameters - conduction and valence band effective densities of states |
| $E_C$, $E_V$, $E_g$ | Band parameters - Conduction and valence band edges and bandgap |

In this study, we limit the interaction to optical and band structure parameters since the device model is only weakly dependent on the other parameters listed. The model structure is a simple inverted structure consisting of electron transport layer, perovskite, and hole transport layer with contacts on a glass substrate, simulated with a transfer matrix methodology and diffusive optics to simulate imperfectly planar surfaces of typical structures. The model outputs include all the usual performance figures of merit as presented in the results section.

## 3 RESULTS AND DISCUSSION

Figure 1 depicts the *Pm3m* cubic unit cell used to perform the calculations and table II gives the results obtained for lattice parameter, band gap and electron affinity, for the cubic $MAPbI_3$: This 96-atoms primitive cell enables us to begin to consider the influence of the distribution of the molecular entity across the lattice on the structural, vibrational and optoelectronic properties of $MAPbI_3$. As noted previously, this unit combined with the optimized Hamiltonian to reproduce the properties of the $MAPbX_3$ perovskites allows the estimation of the lattice parameters and band gaps with an average error of 2. and 5 %, respectively, with respect to the available experimental data. As indicated in the section 2.1, the **PBEx** functional allows us to obtain data of interest in better agreement with experiment than the most commonly used PBE functional (which strongly underestimates the band gap, for instance).

**Table II:** Calculated lattice parameters ($a$ in Å), band gap ($E_g$ in eV) and electron affinity ($\chi$ in eV) for the cubic phase of MAPbI$_3$ at the **PBEx** level. The data obtained at the PBE level (between parentheses) and experimental data are given for comparison.

|   | Calc. | Exp. |
|---|---|---|
| $a$ | 6.368 (6.383) | 6.329[a], 6.308[b] |
| $E_g$ | 1.68 (0.92) | 1.62 (1.50 – 1.69)[c] |
| $\chi$ | 3.79 (4.06) | 3.45[d], 3.90[e], 4.10[f] |

[a]Ref [14]; [b]Ref. [15]; [c]average value of experimental data from Table 2 and between parentheses range of variation of the band gap with different materials formings and measurement techniques cited in Table 1 of Ref. [16], respectively; [d]Ref. [17]; [e]Ref. [18]; [f]Ref. [19].

Table III gives the variation of the band gap and electron affinity for different phases of MAPbI$_3$. For each cell, the systems are fully optimized.

**Table III:** Calculated band gap ($E_g$ in eV) and electron affinity ($\chi$ in eV) for different phases of MAPbI$_3$ at the **PBEx** level. The results on the single cell (12 atoms) are given for comparison.

| Phase | $E_g$ | $\chi$ |
|---|---|---|
| Cubic *Pm3m* | 1.68 | 3.79 |
| Tetra. *I4/mcm* | 2.17 | 3.49 |
| Ortho. *P222$_1$* | 2.25 | 3.46 |
| Single cell | 2.25 | 3.45 |

We note an increase of the band gap and a decrease of the electron affinity with the symmetries lowering of the different phases. As has been noted in the literature (see for instance references 3 and 4), this is due to the combined effects of the octahedra tilting, MA ordering and induced lattice deformations which yields a shift of the top of valence and of the bottom of the conduction bands.

Figure 2 shows the resulting optoelectronic responses for the cubic and tetragonal phase of MAPbI$_3$ compared with experimental data and other DFT calculations at the GW level realized on the single cell [20]. It clearly shows that the proposed method, which takes the MA ordering into account, improves the description of the dielectric responses (notably the peak at 3.5 eV) compared to the local approach based on the single cell or the one obtained at the PBE level. It also illustrates that the phase transitions will directly influence the optical response of the considered perovskites. The corresponding theoretical absorption spectra are in a qualitative agreement with experimental data.

We next evaluated device performance. As mentioned above, the device model simulates standard design consisting of a front ITO surface conductor, MoO$_x$ buffer, PTAA hole transport layer, the perovskite absorber, followed by a SnO$_2$ electron transport layer, the whole on a glass substrate [5].

Table IV gives the performance of devices for different phases of MAPbI$_3$. We note here important advances, which are comparison of **PBEx** functional results and evaluation of the performance of perovskite phases and their stability, a major question in current PSC development.

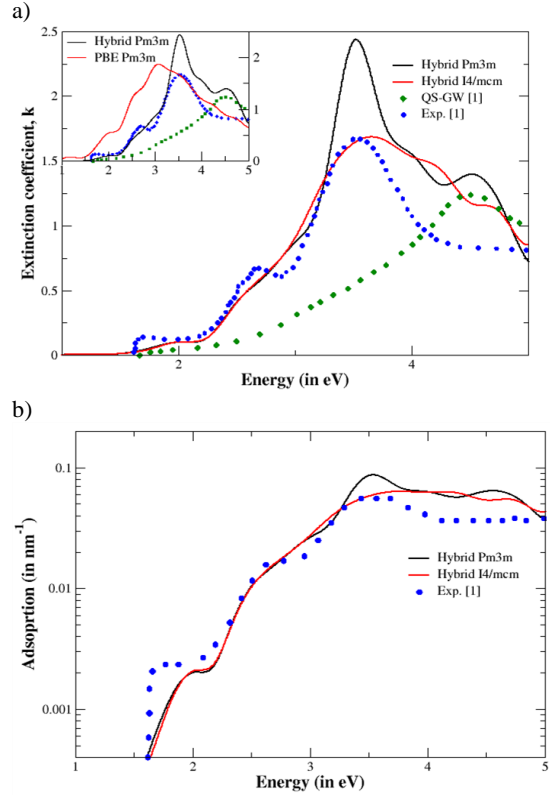

**Figure 2:** a) Obtained extinction coefficient k and b) absorption for the cubic (*Pm3m*, black) and tetragonal (*I4/mcm*, red) phases of MAPbI$_3$. The blue and green dots present the experimental data obtained via ellipsometry on monocrystal and the theoretical results obtained on the pristine cell at the GW level, respectively [20].

**Table IV:** Device performance modelling for parameters taken from the experimental data of literature [20], and from successive theoretical **PBEx** calculations ranging from the single cell (12 atoms cell) to tetragonal and cubic phases (96 atoms cell), showing a peak performance for the cubic phase.

| Data source | Jsc (A/m$^2$) | $V_{oc}$ (V) | $V_{mp}$ (V) | FF (%) | $\eta$ (%) |
|---|---|---|---|---|---|
| Single cell | 13.4 | 1.82 | 1.40 | 71.9 | 17.5 |
| Ortho. *P222$_1$* | 11.6 | 1.80 | 1.42 | 73.6 | 15.5 |
| Tetra. *I4/mcm* | 13.4 | 1.73 | 1.37 | 74.6 | 17.2 |
| Cubic *Pm3m* | 18.2 | 1.25 | 1.11 | 84.3 | 19.2 |
| Exp. [20] | 17.5 | 1.25 | 1.10 | 84.3 | 18.5 |

Following, the example of the dielectric properties, taking into account the MA ordering in MAPbI3 improves the qualitative description of the device performances with respect to experiment. It shows that the best agreement is obtained for the cubic phase of MAPbI3 which possess the highest efficiency. The efficiency of the device decreases with the increase of the bang gap the lowest one corresponding to the orthorhombic phase.

To explain this trend, figure 3 shows band alignments for **PBEx** data set values of affinities and band parameters of the cubic and orthorhombic phases. We note cliffs in absorber-transport layer band profiles (just below 1.2 µm) which translate as drops in charge carriers potential corresponding to drops in maximum power voltage. The significantly greater cliff in the ortho case leads to greater thermalisation losses for both electrons and holes as visible in the step in the electron quasi-Fermi level. This is in part responsible for the lower efficiency of the ortho material compared to the cubic.

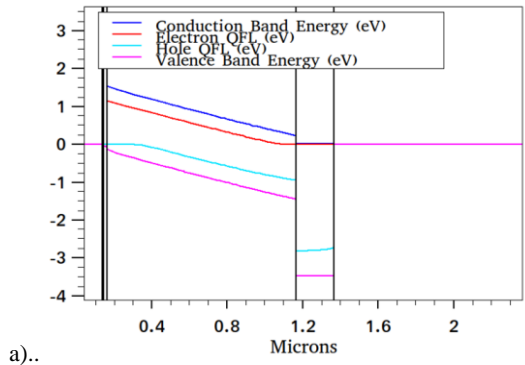
a)..

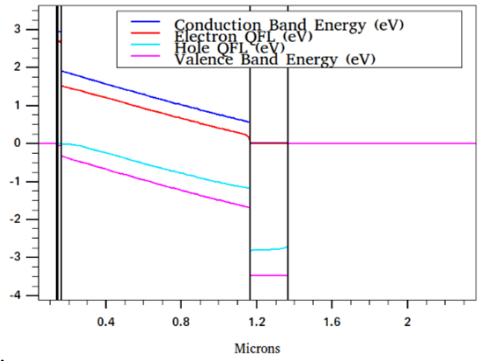
b).

**Figure 3:** Example of a perovskite cell calculated band profile under illumination at short circuit for experimental parameter input values to the device model for a) the cubic and b) the orthorhombic phases. The **PBEx** input to the device model in this case is the 19.2% efficient cubic dataset .

Figure 4 a) shows the corresponding light current curve for which the figures of merit are given in table III. We note here that the lack of steps and flat IV curve for much of the voltage range which corresponds to a high fill factor is evidence of good band alignments in the device. Figure 4 b) shows the (external) quantum efficiency. This shows a broad tail below the electronic gap which is 1.68eV (wavelength 0.74 µm). This requires further work since there is a significant contribution to the photocurrent which is not reflected in experimental data.

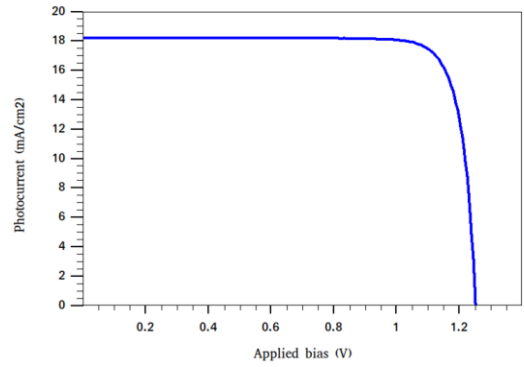
a)

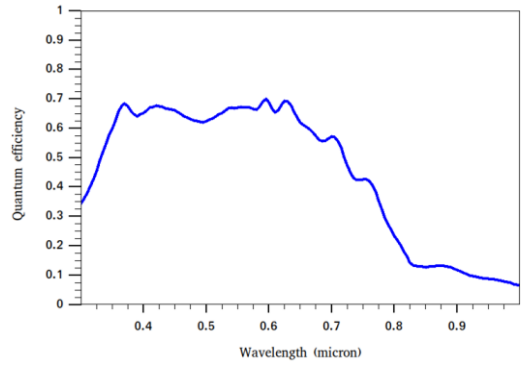
b)

**Figure 4:** Light current (a) and quantum efficiency (b) showing absorption below the gap at 0.74eV which needs further investigation.

## 4 CONCLUSIONS

In conclusion, we present the basis of a pragmatic multiscale approach using atomistic scale first-principles calculations coupled to device scale numerical models. At the first-principles level, a hybrid exchange-correlation functional optimized to yield description of their structural, electronic, and phonon properties in good agreement with experiment, has been used. The obtained band gaps, work functions and dielectric responses served as input data to the device model to estimate the performance of solar cells. The preliminary theoretical atomistic and device model results are both in qualitative agreement with experimental data. This methodology has to be proven on a more detailed sets of perovskites, but, if the trends are confirmed, it might allow to provide a set of criteria for optimizing the materials for different PV applications and for suggesting effective complex perovskites. While the main focus of this work is perovskite materials and therefore of single-junction perovskite solar cells, the extension to tandem solar cells is included given the importance of multijunction device exceeding single junction Shockley-Queisser efficiency limits.

Acknowlegments

The authors thank the ANRT (French National Association for Research and Technology) for its financial support within CIFRE agreement 2023/0728

(industrial convention for training through research), and support from the France 2030 programme PEPR-TASE ("Programme et Equipements Prioritaires de Recherche sur les Technologies Avancées des Systèmes Energétiques") specifically within the MINOTAURE project, Grant ANR-22-PETA-0015.